\documentclass{aa}
\newcommand{\kms}{km s$^{-1}\;$}
\newcommand{\kmss}{km s$^{-1}$}
\newcommand{\km}{km s$^{-1}\;$}

\newcommand{\vlsr}{V$_{\rm LSR}$}
\newcommand{\vs}{V$_{\rm sys}$}

\newcommand{\lsun}{\mbox{L$_{\sun}$}}

\newcommand{\ho}{H$_{2}$O$\;$}
\newcommand{\hho}{H$_{2}$O}
\usepackage{psfig}
\begin{document}
   \authorrunning{Y. Hagiwara et al.}
   \titlerunning{A search for \ho  maser emission towards IRAS galaxies}
%
   \title{A search for extragalactic \ho maser emission towards IRAS galaxies}

   \subtitle{Detection of a maser from an infrared-luminous merger, NGC\,6240}

   \author{Y. Hagiwara,
          \inst{1}
           P. J. Diamond,
           \inst{2}
          \and
           M. Miyoshi
            \inst{3}
          }
   \offprints{Y. Hagiwara}

   \institute{Max-Planck-Institut f\"ur Radioastronomie (MPIfR)
              Auf dem H\"ugel 69, D-53121 Bonn, Germany,
              \email{hagi@mpifr-bonn.mpg.de} 
         \and      
       Jodrell Bank Observatory, University of Manchester,
        Macclesfield, Cheshire, SK11 9DL, UK
         \and
             National Astronomy Observatory,
             Osawa, Mitaka, Japan\\
             }

 \date{Received July 31 2001; accepted November 28 2001}
   \abstract{We report the result of an on-going survey for 22 GHz
   \hho\ maser emission towards infrared luminous galaxies. The
   observed galaxies were selected primarily from the IRAS bright
   galaxy sample. The survey has resulted in the detection of one new
   maser. The new maser was discovered towards the [U]LIRG/merger
   galaxy NGC\,6240, which contains a LINER nucleus. This is the first
   detection of an \ho maser towards this class of galaxy, they are
   traditionally associated with OH megamaser sources. The detected
   maser emission is highly redshifted ($\sim$  260--300 \kmss) with respect
   to the adopted systemic velocity of the galaxy, and we identified
   no other significant emission at velocities $\la$ $\pm$ 500 \km
   relative to the systemic velocity. The presence of high-velocity
   maser emission implies the possible existence of a rotating maser
   disk formed in the merging process. The large maser luminosity
   ($\sim$ 40 \lsun) suggests that an active galactic nucleus could be
   the energy source that gives rise to the water emission.
   Alternatively, the maser emission could be associated with the
   previously observed double radio source in the centre of the
   galaxy. Interferometric observations with high angular resolution
   will be able to clarify the origin of the new maser.
    }
   \maketitle
%
\section{Introduction}
The discovery of the highly Doppler-shifted maser emission at 22 GHz
in the 6$_{16}$--5$_{23}$ transition of \ho in NGC\,4258
(\cite{naka93}), has resulted in a resurgence of single-dish surveys
for extragalactic water masers. Recent maser surveys towards active
galactic nuclei (AGN) have been motivated by the VLBI observations of
a sub-parsec-scale \ho maser disk rotating around a central massive
object in NGC\,4258 (\cite{miyo95}). The detection and imaging of such
\ho $megamaser$ emission enables us to probe the kinematics and
dynamical structures of parsec-scale circumnuclear disks
(\cite{mora99}). \ho megamasers studied to date can be categorised
into two types; $disk~masers$ residing in a (sub-)parsec scale disk
around an active nucleus, and $non-disk~masers$ which are seen
significant distances from the nucleus. The latter can also be
sub-categorised into two types: jet masers associated with radio jets
e.g NGC\,1068 (\cite{gall96}), NGC\,1052 (\cite{clau98}), and Mkn\,348
(\cite{falc00}, \cite{peck01}, \cite{richards01}) or a nuclear
outflowing component such as that observed in the Circinus galaxy
(\cite{linc00}). Masers associated with jets appear to have different
spectral characteristics than the $disk~masers$, the spectra tend to
be broad (few hundred \kmss) and featureless, whereas the $disk~masers$
tend to consist of bright, narrow components clustered in velocity
groups. \ho megamasers generally seem to trace the nuclear activity in
the central parsecs of AGN. To deduce the general properties of such
megamasers and those of the central parsecs of AGN, we need to
increase the number of detections of water megamasers.

\indent At this time, 24 extragalactic water masers inside AGN have
been discovered with single-dish surveys
(e.g. \cite{mora99}). The detection rates of new masers
remain quite low, less than $\sim$ 7\% (e.g., \cite{braa96, braa97,
linc97}). From the presently known sample of sources we have already
deduced some properties of \ho megamasers. They prefer to lie in type
2 Seyfert AGN, suggesting that obscuring gas around an active nucleus
coupled with long gain paths along an edge-on disk in the line of
sight, plays a vital role in giving rise to strong maser emission. A number
of \ho megamasers are known to contain a parsec-scale radio
``core-jet'' structure which the maser can amplify in the gaseous
environment. Accordingly it is reasonable to search for \ho megamasers
towards type 2 Seyfert/LINER nuclei enshrouded by circumnuclear gas
around a compact continuum nucleus.  \\

Here we report the result of an on-going single-dish survey for \ho megamasers towards AGN in infrared (IR)-luminous galaxies, and the discovery of the new maser in the galaxy NGC\,6240.
\section{Sample selection}
\indent The observed galaxies were primarily selected from
far-infrared (FIR) luminous galaxies (IRAS galaxies) with an apparent
FIR luminosity (L$_{\rm FIR}$) $>$ 10$^{11}$ L$_{\odot}$
(\cite{soif89}). Condon et al. (1990) carried out 1.49 GHz imaging
with the VLA at arcsecond resolution for these objects. Due to the now
well-known correlation of L$_{\rm FIR}$ with radio continuum flux
density for the IR-luminous galaxies, Condon et al's work has resulted
in a significant number of detections of radio continuum sources
towards such galaxies. The sample was selected from galaxies in which
the ratio of 1.49 GHz continuum flux density to far-infrared (FIR)
flux density (F$_{\rm 21cm}$/F$_{\rm FIR}$) was $>$ 0.003, where
F$_{\rm FIR}$ = F$_{60\, \mu m}$ + F$_{100\, \mu m}$. About 25 IR-luminous
galaxies were selected in this way, of which 14 sources have been
observed to date. Here we present the initial results of the
survey. The sample observed so far is listed in Table 1.
%
%
%
%
\section{Observations}
The observations of \ho emission (rest frequency: 22.23508 GHz) were
made with the MPIfR 100 m radio telescope at Effelsberg between April
and July 2001, as part of an extragalactic \ho maser survey towards
infrared-luminous galaxies. 15 galaxies have been observed to date
(Table~1). We used a K-band HEMT receiver with two orthogonal linear polarizations. The system temperature was typically 70--110 K,
depending on atmospheric conditions. The amplitude calibration was
based on a flux density at 22 GHz for 3C\,286 (\cite{ott94}), and flux
densities were estimated from the measured antennas temperature,
${T_{\rm A}}^{*}$. The observations were made in position-switching
mode. Typical observing time for each source was about 50--80 minutes,
yielding rms sensitivities of $\sim$ 15--30 mJy. Pointing calibrations
using nearby calibrator sources were carried out during the
observations every $\sim$ 1--1.5 hours. The resultant pointing
accuracy was $<$ 8'' i.e. within 20 \% of the antenna beam size (40''
[FWHM]). Uncertainties in the amplitude calibration are nominally
10--20 \%. The back-end digital autocorrelation system provided 4
autocorrelators for each of the 2 polarisation signals. Each
autocorrelator produced a spectrum in a 40 MHz band (corresponding to
a rest-frame velocity range of 540 \kmss) with 512 spectral channels
providing a frequency resolution of 78 kHz or 1.1 \kmss. The center
frequencies of the four autocorrelators were arranged to have 10 or 20
MHz overlap (depending on source), resulting in a total velocity
coverage of $\pm$ 500 \kms relative to the galaxy's systemic
velocity. After adding the two polarizations together, first or second
order baselines were fitted to, and subtracted from, each of the
spectra.
\section{The survey results}
The survey of 15 sources resulted in the detection of one new \ho
megamaser. On 9 May 2001, we detected \ho emission towards the LINER
galaxy NGC\,6240 (Fig~1a), the detection was confirmed on 14 Jun 2001
(Fig~1b). We searched for water masers in the other 14 AGN but
obtained no detections at a 3 $\sigma$ level of $\sim$ 40--90 mJy
(Table~1). The detection of one new maser among the 15 AGN in the
biased sample yields a detection rate of $\sim$ 6.7 \%, which is
comparable to those of $\sim$7.0 \% in Braatz et al. (1997) and
$\sim$3.4 \% in Greenhill et al. (1997). The \ho maser in the Seyfert
2 galaxy NGC\,5793 was originally selected from this sample, its
discovery stimulated us to continue the programme
(\cite{hagi97,hagi01}). Our survey of IRAS galaxies is still ongoing,
however, if we include the previous detection of the maser in
NGC\,5793, the detection rate in our survey is significantly higher
than those mentioned above. Although the survey is incomplete, its
success seems to partially depend on the assumption that the masing
can be enhanced by ``warm'' dust emission heated by a nuclear radio
source. \\
\section{\ho maser in NGC\,6240}
Fig.1 shows spectra of the \ho maser towards NGC6240 taken with the
100m telescope at Effelsberg. The spectra show a prominent \ho maser
feature with a narrow velocity-width (4.3 \km [FWHM]) centered at
\vlsr = 7565 \km. The feature is redshifted by 261 \km from the
systemic velocity (\vs\ = 7304 \km with respect to the local standard
of rest (LSR), based on HI observations; \cite{deva91}). NGC\,6240
exhibits bright emission with a peak flux density of $\sim$ 40 mJy
(Gaussian-fitted intensity is 36 mJy). Fig.2 shows a spectrum averaged
over the two earlier observing epochs in Fig.1 from which we estimate
the apparent isotropic luminosity of the prominent feature alone to be
$\sim$ 36 L$_{\odot}$ at a distance of 97 Mpc. In addition, we detect
faint narrower (1.1 \km [FWHM]) emission at \vlsr = 7609 \km (3
$\sigma$ level detection). Including both emission features, the total
luminosity of the maser is $\sim$ 40 L$_{\odot}$. Prior to our
observations no \ho emission had been firmly detected
(e.g. \cite{clau86}), although tentative detections had been reported
(\cite{henk84, braa94}). The maser therefore appeared to be in a
flaring state during our observing epochs in 2001. The peak flux
density remained the same within the uncertainties of amplitude
calibrations, at both early observing epochs. However, a third
spectrum, taken on 12 July 2001 shows the strongest feature declining
in flux to $\sim$ 27 mJy while the feature at 7609 \km is more
prominent (Fig~1c).  \\
\indent
NGC\,6240 is categorized as an (ultra) luminous infrared galaxy
([U]LIRG) with a complex structure (\cite{sand88}) and is one of the
most distant megamasers ever detected, rivaling the ``Giga Maser'' in
TXFS\,2226--184, which lies at a distance of 100 Mpc
(\cite{koek95}). The basic characteristics of the galaxy are listed in
Table 2. The large IR luminosity ($> 10^{10.8}$ \lsun) is considered
as having its origin in re-emission from a dust shell surrounding a
hot core, suggesting the presence of an on-going starburst triggered
by the merger of two galaxies (\cite{sand88}). However, recent hard
X-ray measurements detected a strong iron-K emission line at $\sim$6.4
keV with an excess atomic hydrogen column density of $\sim$ 10$^{22}$
cm$^{-2}$ towards the nucleus, a phenomenon which appears to be common
in the presence of an AGN (\cite{iwas98,vign99,ikeb00}). A 5 GHz
high-resolution MERLIN continuum map (Beswick et al. (2001)) revealed
a double source with a separation of $\sim$ 1.5'' ($\sim$ 700
pc). Beswick et al concluded that the radio continuum flux found in
the galaxy is a combination of a starburst and a weak AGN induced by a
merger event. A molecular gas concentration, which lies between the
radio sources, was imaged in CO emission on 100 parsec-scales
(\cite{brya99, tacc99, tecz00}). According to Tacconi et al. (1999),
the CO emission is settling down into the two nuclei (which correspond
to the double peaks in the MERLIN map) and will likely form a central
thin disk in the final stage of evolution.
\section{What is the maser in NGC\,6240?}
The detection of \ho masers from NGC\,6240, which is exhibiting
starburst activity, raises questions concerning the nuclear activity
of the galaxy. The major energy source is considered to be a nuclear
starburst (e.g., \cite{tecz00}) since the ratio of the hard X-ray
luminosity to the IR luminosity is small, $\sim 0.01-0.1$ (Table
2). Such a ratio suggests that the hard X-ray source (i.e. the AGN)
is not dominant (e.g., \cite{ikeb00}). What is the energy source
that gives rise to the water maser in NGC\,6240?

\indent \ho megamasers are often associated with a compact continuum
source such as a radio core or jet. In the nuclear region of NGC\,6240
the two radio sources at 5, 8.4 and 15 GHz remain unresolved at $\sim$
0.1 arcsecond resolution. Each source has a brightness temperature
upper limit (T$_b$) of $\sim 10^4-10^5$ (K) (\cite{colb94,
besw01}). (There has been no reported detection of
milliarcsecond-scale structure in the galaxy.) \\
\indent Tacconi et al. (1999) imaged the CO(2-1) emission peak between
the two radio sources with 0.7'' resolution, the spectral profiles
show wide line-widths spanning $\pm$ 500 \km\ with respect to the
systemic velocity of 7339 \kms. A similar result was obtained in
HCN(1-0) emission, which is a more appropriate tracer of the warm
dense molecular gas that can produce 22\,GHz \ho maser emission. Both
CO and HCN cover the velocities of the maser but there are no distinct
velocity peaks corresponding to those of the maser, weakening the case
for an association between the CO/HCN gas and the masing cloud.

\indent The location of the \ho maser of NGC\,6240 is of great
interest. MERLIN high-resolution HI absorption observations show that
there exist velocity gradients against the two unresolved radio
sources. The HI gas in front of the two sources differ in their
velocity centroids by $\sim 150$ \km; the gradients range from \vlsr =
7100-7200 \km\ and 7250-7350 \km respectively
(\cite{besw01}). However, there is no significant detection of HI gas at
the \ho maser velocities of \vlsr = 7565 and 7609 \kmss, implying the
masers are not associated with the observed HI absorbing gas. It is
thus unlikely that the maser emission is associated with these two
radio sources, which are widely accepted as being supernova remnants.
This is consistent with the fact that total maser luminosity, which we
estimate to be $\sim$ 40 \lsun\ is an order of magnitude larger than
those of extragalactic \ho masers outside AGN (0.1-1.0 \lsun),
i.e. $kilomasers$ that are associated with star forming regions in
nearby galaxies (\cite{clau84, ho87, linc90}). Consequently, it is
likely that the maser in NGC\,6240 is directly associated with nuclear
activity, the presence of which was probed by the hard X-ray
measurements. \\
\indent Given the detection of the features redshifted by 260-300 \km
from \vs\ and the narrowness of each line, it is reasonable to assume
that the maser might originate from a well-defined region where the
masing gas is receding relative to the molecular gas disk observed on
100 pc-scales. One can also speculate on the possible existence of an
edge-on molecular gas disk rotating around a super massive object by
analogy with other megamasers (\cite{mora99}). Such a hypothesis is
supported by the nature of the observed spectrum of the \ho maser
towards NGC6240 and its similarity to that of the so-called
$disk~masers$. If such a scenario was true, the \ho maser should lie
in the dynamical center of the galaxy, where the observed dense
molecular gas will eventually settle down between the two radio
sources and form a disk.\\
\indent NGC\,6240 has often been compared with the prototype ULIRG
Arp\,220 (IC\,4553) from which hard X-ray radiation or other evidence
of AGN activity has never been incontrovertibly detected
(\cite{kii97,genz98,hard98} and references therein). The galaxy is a
merging system showing a large concentration of dense molecular gas
between twin stellar/starburst nuclei (\cite{scov97}). \ho maser
emission has not been detected in Arp\,220 (\cite{braa97}) however,
it is the prototype of the class of galaxies that exhibit OH
megamaser emission. Recent surveys with Arecibo have revealed many new
OH megamaser sources bringing the total number of detections to almost
100 (\cite{darl01}). Most of the OH megamaser detections
are in galaxies that are classified as [U]LIRGs, and all [U]LIRGs
appear to be merging systems. The most remarkable thing is that \ho nuclear megamasers and OH megamasers are mutually exclusive: No galaxy observed to date contains both an OH megamaser and a nuclear \ho megamaser. NGC\,6240 does not show OH emission, however OH gas is seen in intense absorption (e.g. \cite{baan98}). This is similar to the case of the OH absorption towards the nucleus of the megamaser NGC\,5793 (\cite{hagi00}). Until now there have been no detections
of \ho masers towards  merging systems, NGC\,6240 appears to break the
mould.

\indent The presence of strong \ho maser emission and the iron-K line
in NGC\,6240 might mean that the galaxy is experiencing a different
phase of merging from that of Arp\,220 i.e. compact single disk
formation surrounding a ``central engine'', thereby placing it the
final evolution stage in galaxy-galaxy merging.
\section{Conclusions}
We have made the first detection of \ho megamaser emission from a
[U]LIRG/merger galaxy. We have detected emission in the LINER galaxy
NGC\,6240, the maser features are observed to be by 260-300 \km
redshifted with respect to the systemic velocity and to have a
luminosity of ($\sim$ 40 \lsun).  The maser could arise from a dense
circumnuclear molecular cloud on a $\sim 100$ parsec scale or from a
spatially compact maser disk inferred from the presence of the high
velocity maser feature(s). The interpretation of the observed maser is
not unique, though we favour the latter case. Interferometric
observations of the new maser will provide an opportunity to
investigate the kinematics of the ongoing merger.
\begin{acknowledgements}
YH would like to thank H. Imai for his assistance with a part of the
observations. YH also appreciates R. Kawabe for helpful discussion on the source selection, and A. W. Sherwood and Chr. Henkel for able advice during the data analysis. We acknowledge M. J. Claussen for his suggestions on the manuscript. This research has made use of the NASA/IPAC Extragalactic Database (NED) which is operated by the Jet Propulsion Laboratory, California Institute of Technology, under contract with the National Aeronautics and Space
Administration. 
\end{acknowledgements}
\newpage
\onecolumn
\begin{table}
\caption{Observed galaxies}
\label{tbl1}
\begin{tabular}{lllllll}
\hline
%
~~Source &~~$\alpha$$^a$  &~~$\delta$$^a$  &~~\vs~$^{a,b}$  &$\Delta$V$^{c}$ &rms&Epoch  \\ 
&(1950)&(1950)&(\km)&(\km)&(mJy)& \\
\hline
UGC 05101 &09 32 05&+61 34 37&11809&11580--12070&12&A \\
NGC 3310 &10 35 40&+53 45 49&993&500--1520&21&A \\
ARP 299  &11 25 38&+58 51 14&3111&2880--3370&18&C \\  
NGC 3822$^{f}$ &11 39 36&+10 33 19&6120&5340--6610&16&A \\  
NGC 4388 &12 23 15&+12 56 20&2524&2290--2780&28&B\\
NGC 4438 &12 25 14&+13 17 07&71&-160--340&32&B\\
NGC 4490 &12 28 10&+41 55 08 &565&330--820&14&A\\
NGC 4532 &12 31 47 &+06 44 39&2012&1780--2270&25&B\\
NGC 4631 &12 39 42&+32 48 52&606&380--860&12&A \\
NGC 6052 &16 03 01&+20 40 37&4716&4490--4990&14&A\\
NGC 6090 &16 10 24&+52 35 04&8785&8560--9040&25&B\\
{\bf NGC 6240} &16 50 28&+02 28 58&7339& 6850--7870&4.5--7.0&B,C,D\\
NGC 6285 &16 57 37&+59 01 47&5580&5190--6220&14&A\\
NGC 7674 &23 25 24&+08 30 13&8671& 8210--9230&28&A\\
UGC 12914 &23 59 08&+23 12 58&4371&3890--4910&14&A\\
\hline
\end{tabular}
\begin{list}{}{}

\item[$^{\rm{a}}$]{From NED database}
\item[$^{\rm{b}}$]{Heliocentric velocity}
\item[$^{\rm{c}}$]{Observed velocity range}
\item[$^{\rm{d}}$]{Observing epoch; A(23--24 April), B(9--10 May), C(13--14 June), and D(16--17 July) all in 2001}
\item[$^{\rm{e}}$]{Condon et al. 1990}
\item[$^{\rm{f}}$]{The source was not selected from IRAS galaxies}
\end{list}
\end{table}

\begin{table}
\caption[]{NGC\,6240}
\label{tbl-2}
\[
  \begin{array}{p{0.5\linewidth}l}
            \hline
%
Distance$^{a}$                &~~  97 {\rm \; Mpc}  \\ 
Systemic Velocity (21cm HI)$^{b,c}$   &~~  7304\; \pm\; 9\; \rm km\;s^{-1}  \\
~~~~~~~~~~~~~~~~~~~~~~~~~~(optical)$^{b,c}$   &~~  7192\; \pm\; 44\; \rm km\;s^{-1}  \\
Inclination (optical) $^{d}$  &~~60$\degr$ \\
Optical class$^{e}$    & {\rm LINER} \\
F$_{\nu}$ (20~cm)$^{f}$ & 386 \pm 19~{\rm mJy} \\
F$_{\nu}$ (100 $\mu$m)$^{g}$& 27.8 \pm 1.1~{\rm Jy}\\
F$_{\nu}$ (60 $\mu$m)$^{g}$ & 22.7 \pm 0.9~{\rm Jy}\\
F$_{\nu}$ (25 $\mu$m)$^{g}$ & 3.4 \pm 0.1~{\rm Jy}\\
L$_{\rm IR}$ (8 -- 1000 $\mu$m)$^{h}$ &    6.6~10^{11}~\lsun  \\
L$_{\rm X-ray}$ (2 -- 10 keV)$^{i}$ &    4~10^{43}~-~6~10^{44}~{\rm erg~s}^{-1}  \\
\hline

\end{array}
\]
\begin{list}{}{}
\item[$^{\mathrm{a}}$] Assuming $H_0$= 75 km s$^{-1}$ Mpc$^{\;-1}$
\item[$^{\mathrm{b}}$] Converted to LSR from helio centric velocity.
\item[$^{\mathrm{c}}$] de Vaucouleurs et al. (1991)
\item[$^{\mathrm{d}}$] Braatz et al. (1997)
\item[$^{\mathrm{e}}$] Heckman et al. (1987)
\item[$^{\mathrm{f}}$] VLA in B configuration (Colbert et al. 1994)
\item[$^{\mathrm{g}}$] Primary from NED
\item[$^{\mathrm{h}}$] Sanders et al. (1988)
\item[$^{\mathrm{i}}$] Ikebe et al. (2000)
\end{list}
\end{table}

%
\begin{figure}
   \psfig{figure=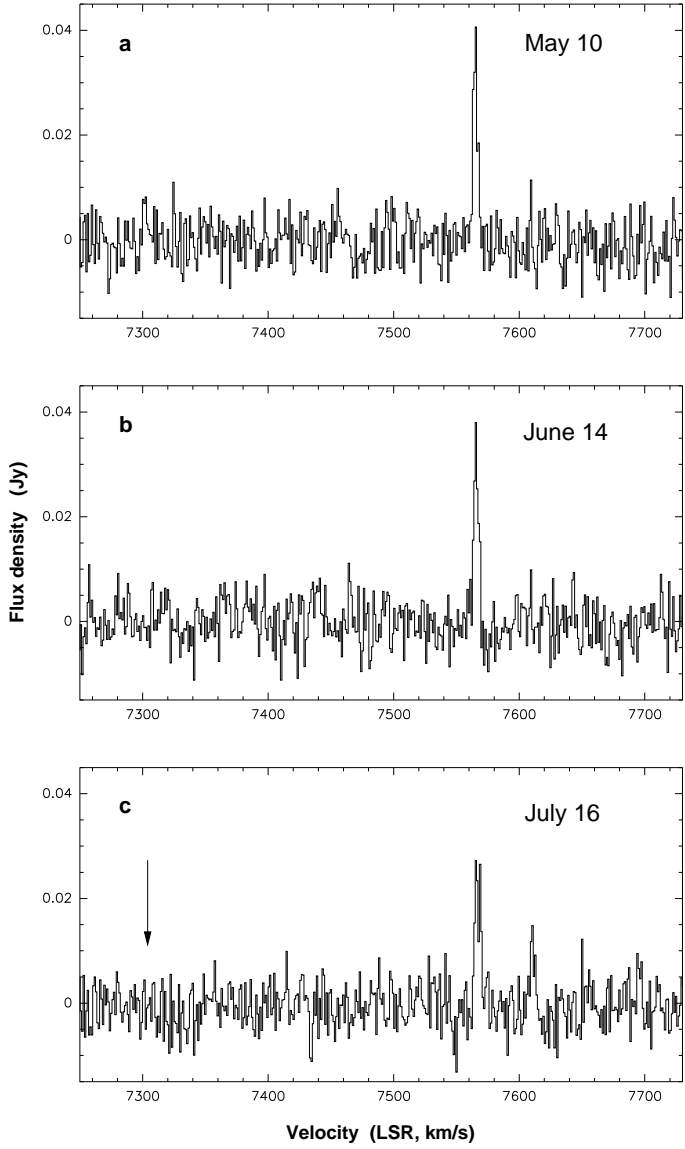,angle=0,width=9cm}
      \caption[]{Spectra of water maser emission towards NGC\,6240 for different three epochs in 2001. The spectral channel spacing is 1.1 \kmss. \vs\ is denoted by an arrow. Note that the flux density of the feature at \vlsr = 7565 \kms clearly decreased at the last epoch (c).}
         \label{fig1}
\end{figure}

\begin{figure}
   \psfig{figure=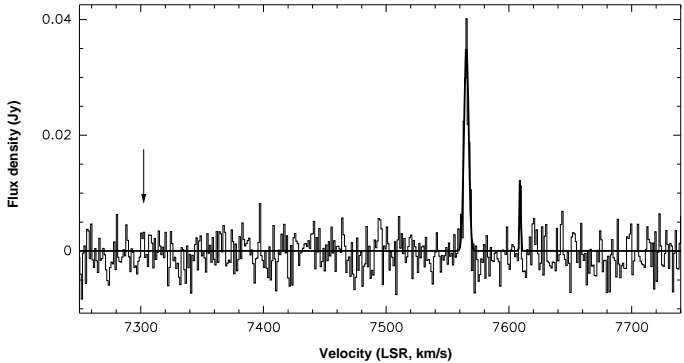,angle=-90,width=9cm}
      \caption[]{Spectrum averaged over two observing epochs on 9 May and 14 June 2001. A downward arrow indicates the adopted systemic velocity of the galaxy, \vlsr=7304 \kmss. Solid line indicates the result of 2-D Gaussian-fitting, assuming two Gaussian components in the observed velocity range. The fitted center velocities of each component are \vlsr=7565 \kmss (FWHM = 4.3 \kms) and 7609 \kms (FWHM = 1.1 \kmss) with peak flux densities of 35 mJy and 15 mJy, respectively.}
         \label{fig2}
\end{figure}
\end{document}